\documentclass[aps,prl,twocolumn,superscriptaddress]{revtex4-2}

\usepackage[T1]{fontenc}
\usepackage{graphicx}
\usepackage{dcolumn}
\usepackage{bm}
\usepackage{amsmath}
\usepackage{braket}
\usepackage[english]{babel}
\usepackage[utf8]{inputenc}
\usepackage[dvipsnames]{xcolor}
\usepackage{soul}
\usepackage{units}
\colorlet{RED}{red}

\usepackage{hyperref}
\allowdisplaybreaks
\begin{document}

\title{A spinless spin qubit}
\author{Maximilian Rimbach-Russ}
\email{m.f.russ@tudelft.nl}
\author{Valentin John}
\author{Barnaby van Straaten}
\author{Stefano Bosco}

\affiliation{QuTech and Kavli Institute of Nanoscience, Delft University of Technology, Delft, The Netherlands
}

\begin{abstract}

All-electrical baseband control of qubits facilitates scaling up quantum processors by removing issues of crosstalk and heat generation. In semiconductor quantum dots, this is enabled by multi-spin qubit encodings, such as the exchange-only qubit, where high-fidelity readout and both single- and two-qubit operations have been demonstrated. However, their performance is limited by unavoidable leakage states that are energetically close to the computational subspace.
In this work, we introduce an alternative, scalable spin qubit architecture that leverages strong  spin-orbit interactions of hole nanostructures for baseband qubit operations while completely eliminating leakage channels and reducing the overall gate overhead. This encoding is intrinsically robust to local variability in hole spin properties and operates with two degenerate states, removing the need for precise calibration and mitigating heat generation from fast signal sources. Finally, our architecture is fully compatible with current technology, utilizing the same initialization, readout, and multi-qubit protocols of state-of-the-art spin-1/2 systems. By addressing critical scalability challenges, our design offers a robust and scalable pathway for semiconductor spin qubit technologies.
\end{abstract}

\date{\today}

\maketitle

\paragraph{Introduction.}
Semiconductor quantum-dot-based spin qubits are a quantum computing platform promising large-scale and fault-tolerant quantum processors using industrial fabrication~\cite{liFlexible300Mm2020,bedecarratsNewFDSOISpin2021,zwerverQubitsMadeAdvanced2022,george12spinqubitArraysFabricated2024,steinacker300MmFoundry2024,huckemannIndustriallyFabricatedSingleelectron2024}. 
Among them, hole spin qubits in silicon (Si) and germanium (Ge) quantum dots are leading candidates for scaling~\cite{lossQuantumComputationQuantum1998,kloeffelProspectsSpinBasedQuantum2013,burkardSemiconductorSpinQubits2023,stanoReviewPerformanceMetrics2022,scappucciGermaniumQuantumInformation2020,fangRecentAdvancesHolespin2023,maurandCMOSSiliconSpin2016,hendrickxSweetspotOperationGermanium2024,borsoiSharedControl162023,johnBichromaticRabiControl2024,zhangUniversalControlFour2024,wangOperatingSemiconductorQuantum2024a}. Their strong spin-orbit interaction (SOI) enables ultrafast all-electric local operations~\cite{bulaevElectricDipoleSpin2007,hendrickxFourqubitGermaniumQuantum2021,froningUltrafastHoleSpin2021,camenzindHoleSpinQubit2022,jirovecSinglettripletHoleSpin2021,wangUltrafastCoherentControl2022,jirovecDynamicsHoleSingletTriplet2022,lilesSinglettripletHolespinQubit2023,liuUltrafastElectricallyTunable2023,wangOperatingSemiconductorQuantum2024a} and long-distance interactions beyond the nearest neighbor~\cite{kloeffelCircuitQEDHolespin2013,mutterNaturalHeavyholeFlopping2021,boscoFullyTunableLongitudinal2022,michalTunableHoleSpinphoton2023,yuStrongCouplingPhoton2023,depalmaStrongHolephotonCoupling2024}. Furthermore, SOI also provides a means to control the properties of the qubit performance in situ, offering sweet spots to improve the performance of the qubit~\cite{mutterAllelectricalControlHole2021,boscoHoleSpinQubits2021,wangOptimalOperationPoints2021,boscoFullyTunableHyperfine2021,piotSingleHoleSpin2022,boscoHoleSpinQubits2022,wangModellingPlanarGermanium2022,senClassificationMagicMagnetic2023,carballidoCompromiseFreeScalingQubit2024,hendrickxSweetspotOperationGermanium2024,mauroGeometryDephasingSweet2024,boscoHighFidelitySpinQubit2024}.
In analogy to superconducting qubits, most spin qubits are controlled using microwave pulses~\cite{ watzingerGermaniumHoleSpin2018,hendrickxFourqubitGermaniumQuantum2021,boscoPhaseDrivingHole2023,saez-mollejoMicrowaveDrivenSinglettriplet2024}, which introduce critical scaling challenges such as cross-talk and heating~\cite{undsethNonlinearResponseCrosstalk2023,undsethHotterEasierUnexpected2023} posing significant barriers to the development of large-scale devices~\cite{veldhorstSiliconCMOSArchitecture2017,vandersypenInterfacingSpinQubits2017}. Progress has been made toward discrete baseband control of hole qubits through spin hopping in Ge quantum dots~\cite{wangOperatingSemiconductorQuantum2024a}, but this approach still requires high-frequency signal generators and high-bandwidth pulses for diabatic gates, adding considerable technological constraints on electronic control.

The exchange-only (XO) qubit does not have such strict electronic requirements. The XO qubit uses multi-spin encoding with degenerate qubit states during idling, making phase tracking and dynamic pulse shaping obsolete and allowing electric two-axis control via inter-dot exchange interactions~\cite{baconUniversalFaultTolerantQuantum2000,divincenzoUniversalQuantumComputation2000,salaExchangeonlySingletonlySpin2017,russThreeelectronSpinQubits2017,burkardSemiconductorSpinQubits2023,boscoExchangeOnlySpinOrbitQubits2024}. XO qubits have shown high-fidelity single- and two-qubit gates, readout, and state initialization~\cite{weinsteinUniversalLogicEncoded2023}. However, XO qubits are heavily impacted by leakage into close-in-energy states, requiring complex pulse sequences for high-fidelity quantum gates and large overhead in fault tolerance. Furthermore, readout and initialization are also limited by additional close-in-energy orbital leakage states~\cite{buterakosSpinValleyQubitDynamics2021}.

\begin{figure}[t!]
    \centering
    \includegraphics[width=\columnwidth]{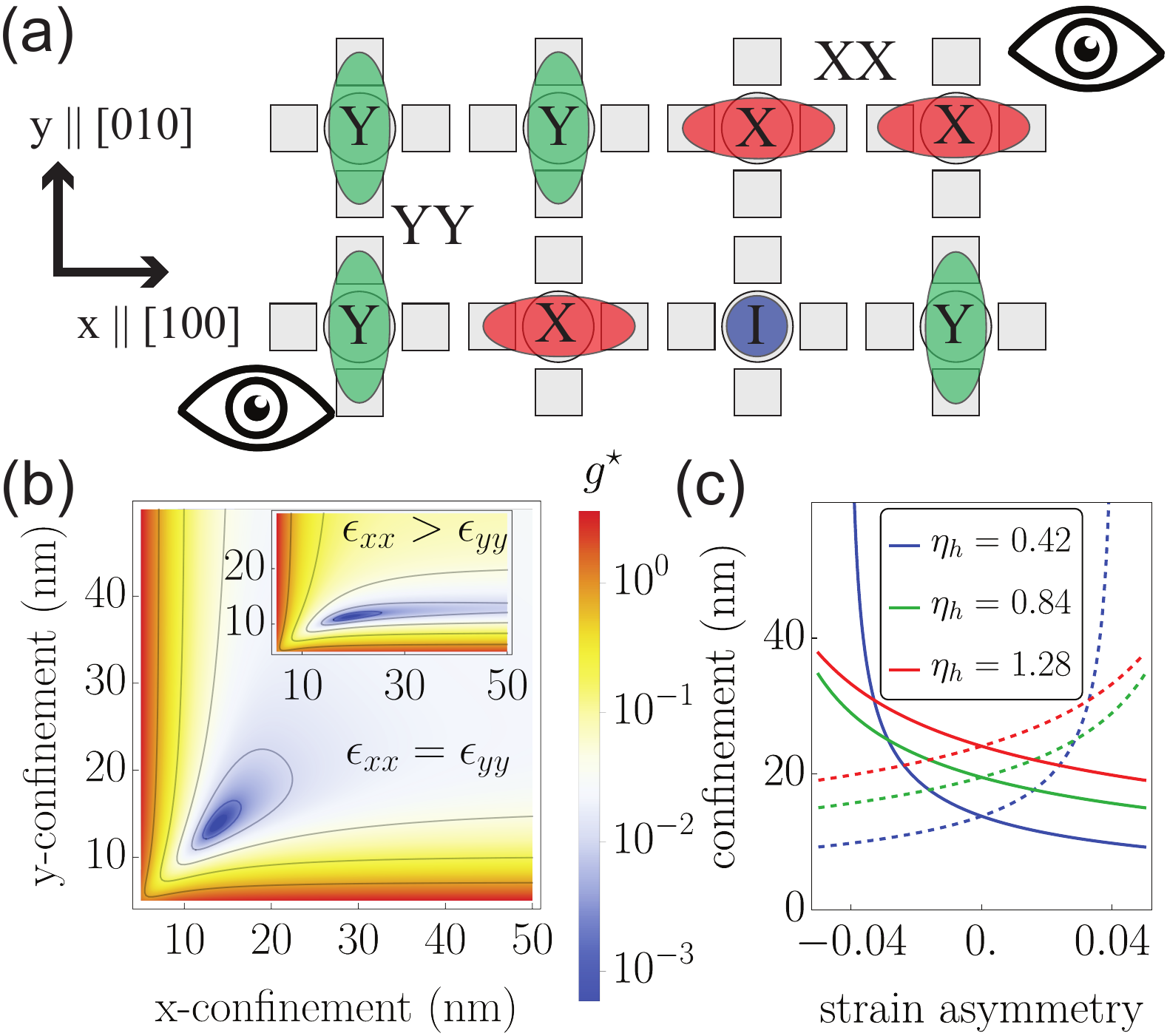}
    \caption{a) Schematics of an architecture for our novel spinless spin qubit (S2) qubit. Each qubit is controlled using 5 gate electrodes. Single-qubit gates are operated by electrically compressing and stretching the qubit. Readout, initialization, and two-qubit gates can be implemented along $x$ and $y$ direction. Readout and initialization uses Pauli-spin-blockade and either requires a close by charge sensor or readout resonator. During idling, the qubit states are not subject to any dynamics. Optimal magnetic field direction is $\boldsymbol{B}\parallel [110]$ or $\boldsymbol{B}\parallel [1\overline{1}0]$ b) Effective g-tensor $g^\star=||\mathcal{G}\boldsymbol{B}||/||\boldsymbol{B}||$ in the optimal direction as a function of characteristic confinement lengths of the groundstate wavefunction in $X,Y$-direction, assuming biaxial strain. The S2 qubit conditions are $\approx\unit[14]{nm}$. The inset shows the conditions assuming non-biaxial strain. c) The S2 qubit conditions for different heavy-hole light-hole coupling strengths as a function of strain asymmetry $(\epsilon_{xx}-\epsilon_{yy})/(\epsilon_{xx}+\epsilon_{yy})$.}
    \label{fig:figure_1}
\end{figure}
In this work, we propose a novel qubit implementation by electrically tuning a hole spin qubit in the presence of strong SOI such that the qubit energy splitting, its Lamor frequency, can be tuned to zero~\footnote{We remark, that artificial spin-orbit interaction through micromagnets, nanomagnets, or magnetic textures can also be used instead~\cite{aldeghiSimulationMeasurementStray2024}.}. Our spinless spin (S2) qubit combines the best features of XO and single spin qubits; simple and fast baseband controlled gates, no phase tracking, simple calibration, and the absence of leakage states. Remarkably, the S2 qubit further enables a scalable architecture (See Fig.~\ref{fig:figure_1}~a)) suppressing additional cross-talk and observer errors, multiple read-out axes, and flexible high-fidelity two-qubit gates.
By mitigating these key issues, our S2 qubit encoding represents paves a path to scalable architectures for a full fault-tolerant quantum processors.

\paragraph{Architecture}
Fig.~\ref{fig:figure_1}~a) shows our envisioned quantum processor architecture. Each qubit requires five gate electrodes for universal control, a central plunger gates surrounded by four control gates, thus, has a similar or smaller gate per qubit count than alternative implementation for baseband controlled qubits~\cite{burkardSemiconductorSpinQubits2023}. We note, that the double barrier design was already demonstrated in Ref.~\cite{borsoiSharedControl162023} and gives rise to a stronger suppression of residual exchange coupling and reduced electrostatic crosstalk due to the larger spacing. By design, such a densely occupied design (with individual control) can be used to implement 2D quantum error corrections codes such as the surface code~\cite{lidarQuantumErrorCorrection2013}. However, we note that using a sparse qubit occupation combined with hopping or shuttling interconnects~\cite{vanriggelen-doelmanCoherentSpinQubit2024,wangOperatingSemiconductorQuantum2024a} or conveying the full qubit~\cite{seidlerConveyormodeSingleelectronShuttling2022,desmetHighfidelitySinglespinShuttling2024}, novel and more efficient low-density-parity-check codes may be implemented~\cite{bravyiHighthresholdLowoverheadFaulttolerant2024,hetenyiTailoringQuantumError2024}.  

\paragraph{S2 Qubit}
For our qubit encoding, we consider a single Kramer doublet of a confined particle in a semiconductor quantum dot. The Hamiltonian of such a system is given by
\begin{align}
    H=\frac{1}{2}\mu_B \boldsymbol{\sigma}\cdot \mathcal{G}\boldsymbol{B}.
\end{align}
Here, $B=(B_x,B_y,B_z)^T$ is the magnetic field vector, $\mu_B$ is Bohr's magneton, and $\mathcal{G}$ is the (anisotropic) gyro-magnetic matrix $3\times 3$~\cite{abragamElectronParamagneticResonance2012}, also called the g-tensor, of the doublet. For our qubit design, we use two properties. Firstly, at least two eigenvalues of the g-tensor, so-called principal values, can be electrically tuned to zero. Secondly, the magnetic field is not collinear with one of the principal axes of the g-tensor and is orthogonal to the principal axis with non-zero eigenvalue. While not exclusive, these requirements are often found in semiconductor hole qubits. Examples, that fulfill this condition are silicon and germanium heterostructures~\cite{boscoSqueezedHoleSpin2021,scappucciGermaniumQuantumInformation2020,delvecchioLightholeGatedefinedSpinorbit2023}. 

Without loss of generality, we now consider for the reminder of the article planar germanium grown in [001] directions and prove that both requirements can be fulfilled in state-of-the-art materials. The g-tensor for planar strained germanium approximately consists of three parts~\cite{abadillo-urielHoleSpinDrivingStrainInduced2023,supplemental_material} $\mathcal{G}=\mathcal{G}_0+\mathcal{G}_\text{strain}+\mathcal{G}_\text{conf}$. The first term is the bare g-tensor, which for heavy-hole ground states is given by $\mathcal{G}_0=\text{diag}(3q,-3q,6\kappa +27q/2)$ and is highly anisotropic $|\kappa| \gg |q|$. The second and third expressions are g-tensor corrections arising from non-axial strain components and from confinement.

Explicitly, considering an in-plane magnetic field, the Hamiltonian of the ground doublet is given by
\begin{align}
    H=& \frac{1}{2}\mu_B (B_x \mathcal{G}_{xx}+B_y \mathcal{G}_{xy}) \sigma_x + \frac{1}{2}\mu_B (B_y \mathcal{G}_{yy}+B_x \mathcal{G}_{yx}) \sigma_y
    \label{ham:Zeeman}
\end{align}
with the electrically tunable g-tensor components~\cite{venitucciSimpleModelElectrical2019,michalLongitudinalTransverseElectric2021,martinezHoleSpinManipulation2022a,abadillo-urielHoleSpinDrivingStrainInduced2023,supplemental_material}
\begin{align}
    \mathcal{G}_{xx} &\approx 3q - \frac{6 \widetilde{\kappa} b_v (\braket{\epsilon_{xx}}-\braket{\epsilon_{yy}})}{\Delta_\text{HL}} - \frac{6 (\lambda \braket{p_x^2}-\lambda^\prime\braket{p_y^2})}{m_0 \Delta_\text{HL}}\\
    \mathcal{G}_{yy} &\approx -3q - \frac{6 \widetilde{\kappa} b_v (\braket{\epsilon_{xx}}-\braket{\epsilon_{yy}})}{\Delta_\text{HL}} + \frac{6 (\lambda \braket{p_y^2}-\lambda^\prime\braket{p_x^2})}{m_0 \Delta_\text{HL}},\\
    \mathcal{G}_{xy,yx} &\approx \pm  \frac{4\sqrt{3} \kappa d_v \epsilon_{xy}}{\Delta_\text{HL}} \mp \frac{12 \widetilde{\lambda} \braket{p_x p_y}}{m_0 \Delta_\text{HL}},
\end{align}
Here the average $\braket{\cdot}$ is with respect to the in-plane component of the groundstate wavefunction. The material dependent scaling parameters $\lambda= 2\eta_h \gamma_3^2-\widetilde{\kappa} \gamma_2$, $\lambda^\prime= 2\eta_h \gamma_3 \gamma_2-\widetilde{\kappa} \gamma_2$, $\widetilde{\lambda}= 2\eta_h (\gamma_2 \gamma_3+\gamma_3^2)-\widetilde{\kappa} \gamma_3$, and $\widetilde{\kappa}=\kappa-2\widetilde{\eta}_h\gamma_3$ depend on the material-dependent Luttinger parameters $\gamma_1$, $\gamma_2$, and $\gamma_3$, the material dependent deformation parameters $a_v$, $b_v$, and $d_v$~\cite{terrazosTheoryHolespinQubits2021}, and the inter-band couplings, $\eta_h$ and $\widetilde{\eta}_h$, that depend on matrix elements between out-of-plane wave functions. Conveniently, $\eta_h$ and $\widetilde{\eta}_h$ and $\braket{p_i p_j}$ can be electrically controlled. In principle, $\braket{\epsilon_{ij}}$ is also electrically tunable but strain is in state-of-the-art devices less predictable.
In Figure~\ref{fig:figure_1}~b) we show the energy splitting of Hamiltonian~\eqref{ham:Zeeman} considering negligible shear strain, $\braket{\epsilon_{xy}}=\braket{\epsilon_{xz}}=\braket{\epsilon_{yz}}=0$ and separable wave-functions $\braket{p_x p_y}=\braket{p_x p_z}=\braket{p_x p_z}=0$ with and without axial-symmetric strain. Here, we introduced for better understanding the characteristic length of the in-plane wavefunction $\braket{p^2_i}=\hbar^2/(2L_i^2)$ with $i=x,y$. We can clearly identify the operation regime of our S2 qubit at characteristic confinement lengths of $\unit[10-20]{nm}$ using parameters from Ref.~\cite{abadillo-urielHoleSpinDrivingStrainInduced2023}. The zero-crossings can be analytically expressed as 
\begin{align}
    \braket{p_x^2} &= p_{x,a}^2 = \frac{m_0}{2} \left[\frac{\Delta_\text{HL} q}{\lambda-\lambda^\prime } - \frac{2 b_v \widetilde{\kappa}  (\braket{\epsilon_{xx}}-\braket{\epsilon_{yy}})}{\lambda + \lambda^\prime}\right] \\
    \braket{p_y^2} &= p_{y,a}^2 = \frac{m_0}{2} \left[\frac{\Delta_\text{HL} q}{\lambda-\lambda^\prime } + \frac{2 b_v \widetilde{\kappa}  (\braket{\epsilon_{xx}}-\braket{\epsilon_{yy}})}{\lambda + \lambda^\prime}\right].
\end{align}
While the required characteristic confinement lengths seem small, simulations and experimental demonstration show that they are feasible~\footnote{Supplement Fig.~S22~b) in Ref.~\cite{wangOperatingSemiconductorQuantum2024a} }. In Fig.~\ref{fig:figure_1}-c), we show the S2 condition for varying strain asymmetry $(\epsilon_{xx}-\epsilon_{yy})/(\epsilon_{xx}+\epsilon_{yy})$ that may arise from disorder~\cite{corley-wiciakNanoscaleMapping3D2023} for different heavy-hole light-hole coupling strengths. Conveniently, typical strain fluctuations arising from dislocations are on a length-scale multiple times the size of single qubits~\cite{corley-wiciakNanoscaleMapping3D2023}, thus each S2 qubit can be tuned to its own respective operation point. We also remark, that the S2 condition is predicted to be possible in different material and device structures~\cite{boscoSqueezedHoleSpin2021,boscoFullyTunableHyperfine2021,delvecchioLightholeGatedefinedSpinorbit2023}. 

Additionally, shear strain from metallic top gates~\cite{abadillo-urielHoleSpinDrivingStrainInduced2023} or strong confinement~\cite{martinezHoleSpinManipulation2022} adds terms of the form $(B_x \mathcal{G}_{zx}+B_y \mathcal{G}_{zy})\sigma_z$ to  Hamiltonian~\eqref{ham:Zeeman} giving rise to errors. However, due to the interplay of confinement and strain terms, there exist a confinement that fulfills our requirements (see general expressions in Ref.~\cite{supplemental_material}). In practice, one wants to avoid these shear contributions, for example, through deeper quantum wells, gate designs that minimize average shear strain, or materials with a smaller differential thermal expansion coefficient.

\paragraph{Qubit states}
The qubit states are encoded in the two-fold degenerate Kramer doublet, defined as $\ket{\uparrow,\downarrow}$. Unlike conventional Loss-DiVincenzo qubits and similar to exchange-only qubits, the degeneracy ensures that our qubit has no dynamics during idling. Unlike exchange-only qubits, however, our qubit has no leakage states since only a single Kramer doublet is used and excited states are far-off separated by more than 1 meV. Therefore, our design combines the best of both worlds. To the best of our knowledge, there is no equivalent encoding in any quantum computing platform. 

\begin{figure}[t]
    \centering
    \includegraphics[width=\columnwidth]{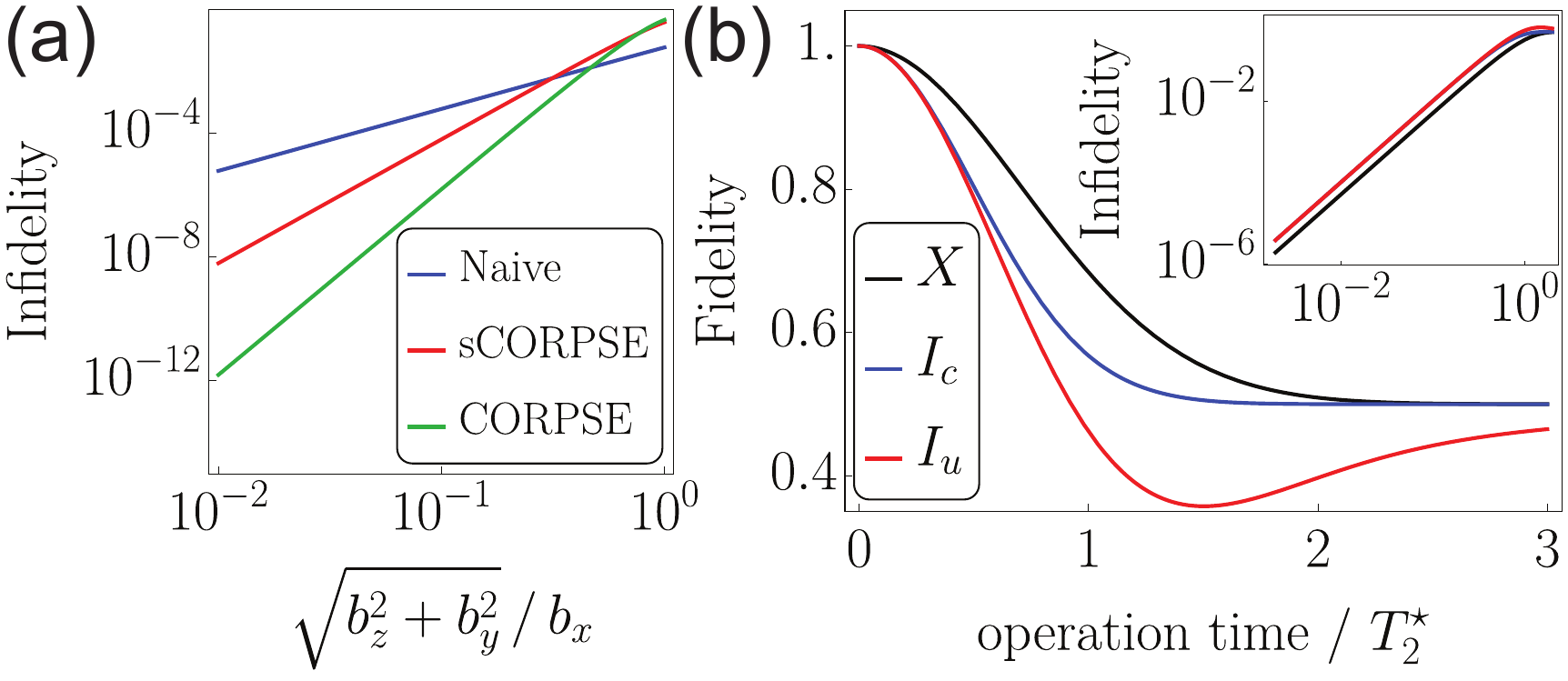}
    \caption{a) Single qubit gate fidelity~\cite{fidelity} of a $X$-gate as a function of miscalibration $b_{y,z}\neq 0$ from the S2 condition without decoherence effects. Three pulse techniques are shown, using a (blue) single pulse, (red) composite pulse sequence short-CORPSE, and (green) composite pulse sequence CORPSE. b) Fidelity of a (black) $X$-operation and (blue and red) idling $I$-operation under the influence of low frequency noise coupling in via the g-tensor as a function of operation time. Our simulation assumes fluctuations in $b_x$ and $b_y$, uses the quasistatic approximation, neglects transversally coupled noise for the $X$-operation, and in the case of idling, (blue) correlated and (red) uncorrelated fluctuations between $b_x$ and $b_y$. }
    \label{fig:figure_2}
\end{figure}

\paragraph{Single qubit operations}
Single qubit gates are enabled by small deviations from the degeneracy condition $\braket{p_x^2}\rightarrow \braket{p_x^2} + p_{x,a}^2$ and $\braket{p_y^2}\rightarrow \braket{p_y^2} + p_{y,a}^2$ that lead to the standard Hamiltonian
\begin{align}
    H_{1q} &= b_x  \sigma_x + b_y \sigma_y
    \label{eq:QubitHam}
\end{align}
with $b_{x,y} = -\frac{1}{2}\mu_B B_{x,y} \frac{3}{m_0 \Delta_\text{HL}} (\lambda \braket{p_{x,y}^2}-\lambda^\prime\braket{p_{y,x}^2})$ and 
neglecting out-of-plane tilts. 

Using for example the layout  shown in Fig.~\ref{fig:figure_1}~a), $\braket{p_x^2}$ and $\braket{p_y^2}$ can be individually and electrically controlled, giving rise to arbitrary single qubit rotations. Orthogonal rotations around the $X,Y$-axis can be realized by pulsing $\frac{\braket{p_{x,y}^2}}{\braket{p_{y,x}^2}}=\frac{\lambda}{\lambda^\prime}$~\footnote{Alternatively, rotations around the $\sigma_x\pm\sigma_y$-axes can be achieved by the condition $\braket{p_y^2}(\beta^2\lambda-\lambda^\prime)=(\beta^2\lambda^\prime-\lambda)\braket{p_x^2}$ with $\beta=B_y/B_x$.}. Notably, a misaligned magnetic field only affects the gate speed and not the rotation angle of the quantum gate. Consequently, the fastest operations are achieved if $\boldsymbol{B}\parallel [110]$ or $\boldsymbol{B}\parallel [1\overline{1}0]$. We remark that in realistic structures due to cross-capacitance coupling, orthogonal control might require an additional calibration step~\cite{martinsNoiseSuppressionUsing2016} or sufficient virtualization~\cite{hsiaoEfficientOrthogonalControl2020}. 

Potential error sources for the gate operations are additional terms in Hamiltonian~\eqref{eq:QubitHam} proportional to $\sigma_z$ and calibration errors that lead to $b_{y,x}\neq 0$ while doing $X,Y$-operations. Fortunately, there are several techniques being able to suppress such errors. For small to medium-size $\sigma_z$-terms, composite pulse sequences can be used to drastically suppress off-resonant errors~\cite{cumminsTacklingSystematicErrors2003}. In Fig.~\ref{fig:figure_2}-a), we show the infidelity of $X$-operations using naive direct pules, the composite pulse short-CORPSE (sCORPSE), and the composite pulse sequence CORPSE. Conveniently, the CORPSE pulse sequence for a $X,Y$-pulse only requires $X,Y$ and $-X,-Y$ pulses, thus no additional constraint on the electronics is imposed and no additional calibration is needed. For large $\sigma_z$ terms, non-adiabatic gates can be used instead~\cite{wangOperatingSemiconductorQuantum2024a}. We further note, that due to the degeneracy, Holonomic single-qubit gates might be possible~\cite{zhangGeometricHolonomicQuantum2023,kolokProtocolsMeasureNonAbelian2024}.

Analogously to almost all qubit encodings, the S2 qubit is also subject to electric and magnetic noise. Electric noise dominantly arises from ubiquitous charge noise present in condensed matter that couples to the electrostatic confinement and wavefunction of the charge carrier and affects the qubit via spin-orbit and exchange couplings. Unlike qubits with a finite frequency splitting during idling, zero-frequency qubits are subject to a different type of decoherence in the presence of noise, and distinction of the terms $T_1$ and $T_2$ is obsolete. To still provide a meaningful description, we instead compute and report the process fidelity of the operations~\cite{fidelity}, including the idling operation, as a function of time. By design, decoherence during an $X,Y$-operation corresponds to the $T_2^\star$-time of conventional spin-1/2 qubits. However, note that the gate time of performing $X,Y$-operations is given by the change in Lamor frequency instead of the Rabi frequency, which can be multiple times faster. Quality factors $f_\text{Lamor}\times T_2^\star$ exceeding 100 are frequently reported~\cite{stanoReviewPerformanceMetrics2022,hendrickxSweetspotOperationGermanium2024,wangOperatingSemiconductorQuantum2024a}. Using Schrödinger-Poisson simulations~\cite{philippopoulosAnalysis3DTCAD2024}, we estimate a central gate sensitivity $\partial_{V_c}\braket{p_{y}^2}/\hbar^2 = \unit[(0.064)^2]{nm^{-2}/V}$ giving rise to $b_{x}\approx\unit[630]{\frac{MHz}{V}}$ considering $B=\unit[100]{mT}$ and $\eta=0.42$. Faster operations can be achieved by pulsing center and side gates simultaneously. Further improvement in gate fidelity can be gained by more sophisticated composite pulse sequence design~\cite{ichikawaDesigningRobustUnitary2011}. 

Decoherence during idling is caused by two orthogonally coupled random fluctuations and is approximately $\sqrt{2}$-times faster depending on their cross-correlation. In Fig.~\ref{fig:figure_2}~b) we show the gate fidelity as a function of operation time and $T_2^\star$ for single-qubit gates $X$ and the idling operation. Using a lower estimate of $Q=100$, gate infidelities above $1-F=10^{-4}$ are achievable for all operations. Since charge and magnetic noise is dominantly slow, dynamical decoupling protocols, such as echo using Z pulses or derivates of the XY4 protocol~\cite{gullionNewCompensatedCarrPurcell1990,wangEffectPulseError2012,wangComparisonDynamicalDecoupling2012}, can greatly extend the coherence time. Conveniently, the XY4 protocol and its derivates also decouples pulse errors.

Magnetic noise typically originates from nuclear spins with non-zero spin that are coupled to the spin directly through the hyperfine tensor. For planar germanium, the hyperfine tensor is highly anisotropic, thus, magnetic noise is highly suppressed for in-plane magnetic field directions.  Remarkably, the S2 operation regime coincides in some materials with the noise sweet spot~\cite{boscoHoleSpinQubits2021,boscoFullyTunableHyperfine2021}. Furthermore, the nuclear spins as a magnetic noise source can be effectively eliminated by isotopic enrichment of spinless isotopes.

\begin{figure*}[t]
    \centering
    \includegraphics[width=1.9\columnwidth]{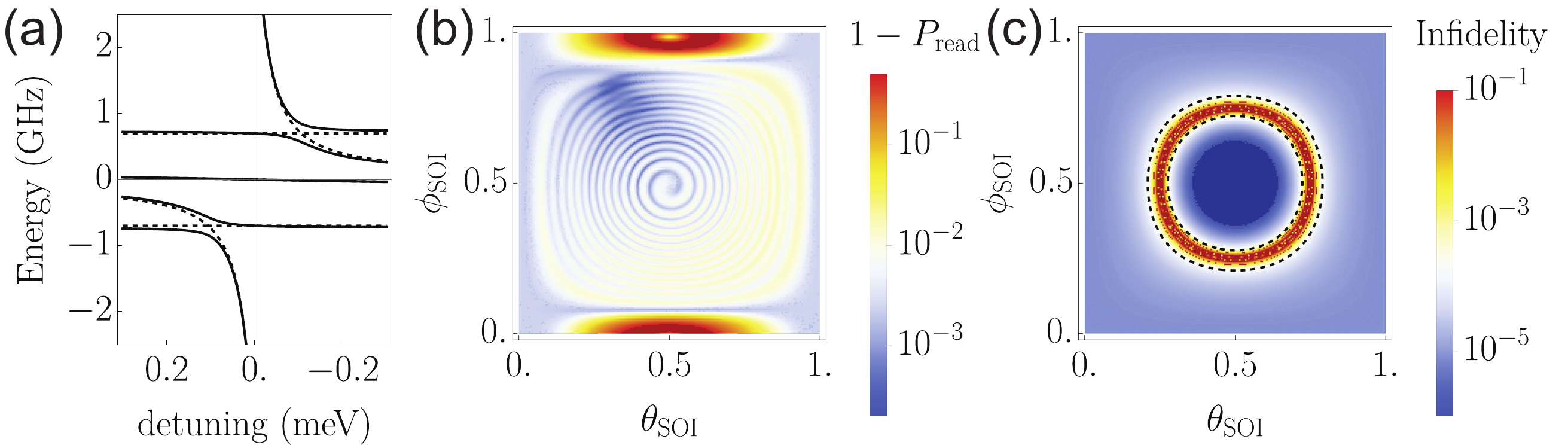}
    \caption{a) Energy diagram of the anti-crossing used for Pauli-Spin Blockade (PSB) readout as a function of detuning $\varepsilon$ for stabilized protocol $b_y^{(A)}=-b_y^{(B)}=\unit[0.1]{GHz}$ (dashed lines showing $b_y^{(A)}=b_y^{(B)}=0$) considering $b_x^{(A)}=\unit[0.3]{GHz}$, $b_x^{(B)}=\unit[0.4]{GHz}$, $t_c=\unit[3]{GHz}$, and $\theta_\text{SOI}= 0$. Our protocol induces an anti-crossing between $\ket{00}$ and $\ket{S_{02}}$ identical as in the presence of finite spin-orbit interaction. b) Simulated readout infidelity of the stabilized protocol for a linear ramp from $\varepsilon(t=0)=\unit[0.5]{meV}$ to $\varepsilon(t=T)=-\unit[0.5]{meV}$ with duration $T=\unit[52.7]{ns}$ as a function of the spin-flip angle $\theta_\text{SOI}$ and SOI vector $\boldsymbol{n}=(\cos(\Phi_\text{SOI}),\sin(\Phi_\text{SOI}),0)^T$ and using the same parameters as before. c) Simulated average gate infidelity of an adiabatic CX two-qubit gate up to single-qubit X gates with fixed duration $T=\unit[36]{ns}$ using a Tukey pulse~\cite{rimbach-russSimpleFrameworkSystematic2023} as a function of the spin-flip angle $\theta_\text{SOI}$ and SOI vector $\boldsymbol{n}=(\cos(\Phi_\text{SOI}),\sin(\Phi_\text{SOI}),0)^T$. The dashed black lines highlight the problematic regime in which SWAP oscillations can be hardly suppressed~\cite{supplemental_material}. Simulation parameters are identical to a) and b). Tunneling and exchange are modelled assuming Gaussian overlaps~\cite{burkardCoupledQuantumDots1999}. }
    \label{fig:figure_3}
\end{figure*}

\paragraph{Readout and initialization}

We read out our qubit using conventional Pauli-spin blockade (PSB) methods, where spin-selective charge transitions are observed. While PSB readout is generally used to read out any quantum dot spin-qubit, the spin state of the unblocked state depends on the interplay of the Zeeman interaction and the exchange interaction between spin in neighboring quantum dots. We distinguish three cases. i) For a uniform magnetic field and a uniform g-tensor without spin-orbit interaction, the spin singlet state $\ket{S}=(\ket{10}-\ket{01})/\sqrt{2}$ is unblocked. We call this PSB-S and is typically used for electron spin qubits without micromagnet, singlet-triplet, and exchange-only qubits. ii) For magnetic fields with a parallel gradient or a uniform magnetic field along a (common) principal axis of the g-tensors without spin-orbit interaction, the $\ket{01}$ or $\ket{10}$ state is unblocked. We call this the PSB-Z and is typically used for electron spin qubits with a micromagnet. iii) For magnetic fields with an orthogonal gradient, a uniform magnetic field along a (common) principal axis of the g-tensors, or spin-orbit interaction, the $\ket{11}$ is unblocked. We call this the PSB-T and is typically implemented for hole spin qubits.

The Hamiltonian for readout can be well approximated by~\cite{danonPauliSpinBlockade2009}
\begin{align}
    H=& H_{1q}^{(A)}+H_{1q}^{(B)} + \varepsilon \ket{S_{02}}\bra{S_{02}}+\sqrt{2}t_c\cos(\theta_\text{SOI})\ket{S}\bra{S_{02}} \nonumber\\
    &+i\sqrt{2}t_c\sin(\theta_\text{SOI})\,\boldsymbol{n}\cdot\ket{\boldsymbol{T}}\bra{S_{02}} +\text{h.c.},
\end{align}
where $H_{1q}^{(A,B)}$ is the single-qubit Hamiltonian~\eqref{eq:QubitHam} of qubit $A,B$, $\boldsymbol{n}$ is the normalized spin-orbit vector, $\theta_\text{SOI}$ the spin-flip angle, and $\boldsymbol{T}_{AB}=(T_x,T_y,T_z)^T$ is a vector consisting of the spin-1 triplet states with $\ket{T_{x,y}}=i^{1/2\mp 1/2}(\ket{00}\mp\ket{11})/\sqrt{2}$ and $\ket{T_z}=(\ket{01}+\ket{10})/\sqrt{2}$. For hole implementations in planar heterostructure, $\theta_\text{SOI}$ arises from cubic-in-momentum spin-orbit coupling~\cite{marcellinaSpinorbitInteractionsInversionasymmetric2017} or linear-in-momentum from gradients in strain~\cite{abadillo-urielHoleSpinDrivingStrainInduced2023}, and can be assumed or engineered to be small~\cite{lodariLightlyStrainedGermanium2022,stehouwerGermaniumWafersStrained2023,frinkReducingStrainFluctuations2024}. For other platforms, however, $\theta_\text{SOI}$ can be size-able~\cite{scappucciGermaniumQuantumInformation2020}.
Remarkably, our qubit can be electrically tuned into all mentioned PSB readout types if the spin-flip angle $\theta_\text{SOI}\approx m\pi $ with integer $m$. For $b_x^A=b_x^B=b_y^A=b_y^B$ during the charge transition, we find PSB-S, for $b_{x,y}^A\neq b_{x,y}^B\neq 0$ and $b_{y,x}^A=b_{y,x}^B=0$ we find PSB-X,Y which is PSB-Z in the x,y-basis, and for $b_{x,y}^A\neq b_{y,x}^B\neq 0$ we find PSB-T readout. 

To avoid this variability, we conceive a two-step PSB readout protocol for high-fidelity readout and initialization: i) We elongate the wave-functions such that they get closer and the tunnel coupling between the two dots is increased. For PSB-X, this leads to $b_{x}^{A,B}>0$. ii) While tilting the two dots by sweeping the detuning $\varepsilon$ from positive to negative, we additionally compress (stretch) the first (second) wavefunction such that $b_{y}^{A}\approx - b_{y}^{B}$~\footnote{In practice, maintaining the condition $b_{y}^{A}=b_{y}^{B}=0$ during the readout process might be challenging. The only strict requirement for our protocol is to first turn on one axis and then maintain adiabaticity while turning on the other parameter.}. The second step ensures that we open the anti-crossing between the between $\ket{00}$ and $\ket{S_{02}}$ even in the presence of small $\theta_\text{SOI}$ and ensure PSB-T (see Fig.~\ref{fig:figure_3}~a)). Fig.~\ref{fig:figure_3}~b) shows the readout infidelity $1-P_\text{read}$ for a constant detuning pulse, where $P_\text{read}=P_{\ket{00}\rightarrow \ket{S_{02}}}-P_{\ket{10}\rightarrow \ket{S_{02}}}$ is the readout visibility~\cite{seedhousePauliBlockadeSilicon2021}. Noticeably, high-fidelity readout is achieved for the vast majority of conditions. Only for $\theta_\text{SOI}\approx \pi/2$ and $\phi_\text{SOI}\approx 0$ the anti-crossing between $\ket{00}$ and $\ket{S_{02}}$ is closed due to the interplay between spin-orbit interaction and quantization axis rotation~\cite{jirovecDynamicsHoleSingletTriplet2022}. The oscillations originate from Landau-Zener transitions. We further remark, that faster pulses with higher fidelity can be achieved using optimal control, especially the QUAD protocol allows for consistently high transfer fidelity~\cite{fernandez-fernandezQuantumControlHole2022,fehseGeneralizedFastQuasiadiabatic2023,meinersenQuantumGeometricProtocols2024}.  
Consequently, high-fidelity state initialization can be achieved by reversing the upper protocol.

\paragraph{Two qubit gates}
The two qubit gates are implemented similar to other spin qubit implementations using the exchange interaction originating from the overlap of the wave-function and the Coulomb interaction. The interaction Hamiltonian between qubit A and B can be approximated by~\cite{geyerAnisotropicExchangeInteraction2024,zhangUniversalControlFour2023,saez-mollejoMicrowaveDrivenSinglettriplet2024} 
\begin{align}
    H_{2q}= H_{1q}^{(A)}+H_{1q}^{(B)} + \frac{J}{4} \boldsymbol{\sigma}^{(A)}\cdot R_{\boldsymbol{n}} (2\theta_\text{SOI})\boldsymbol{\sigma}^{(B)}
\end{align}
where $J$ is the electrically tunable exchange coupling and $R_{\boldsymbol{n}} (2\theta_\text{SOI})$ is a 3D rotation matrix originating from the influence of the spin-orbit interaction~\cite{geyerAnisotropicExchangeInteraction2024}. In our proposed architecture in Fig.~\ref{fig:figure_1}~a), two-qubit gates can be implemented in the $X,Y$ direction, giving rise to a maximally entangling \textsc{cnot} gate up to single-qubit rotations. Additionally, the two-qubit gate can be implemented using adiabatic or diabatic pulses~\cite{burkardPhysicalOptimizationQuantum1999,meunierEfficientControlledphaseGate2011,rimbach-russSimpleFrameworkSystematic2023}. An undesired evolution during the diabatic gate can be strongly suppressed by interleaving the two-qubit gate with single-qubit $X,Y$ gates~\cite{lossQuantumComputationQuantum1998,russHighfidelityQuantumGates2018}. Fig.~\ref{fig:figure_3}~c) shows the average gate fidelity~\cite{fidelity} for an adiabatic \textsc{cnot} gate in the $x$-direction for a maximum qubit frequency difference of $\Delta E=\unit[200]{MHz}$. Noticeably, there are three regimes showing vastly different infidelities. The outer area is the conventional adiabatic regime, where SWAP oscillations giving rise to non-adiabatic dynamics are sufficiently suppressed. In the area highlighted by the dashed lines~\cite{supplemental_material}, the adiabatic condition is not fulfilled as the entangling phase becomes very small, giving rise to large diabatic errors. In the center area, the diabatic contributions are strongly suppressed, giving rise to the highest fidelities~\cite{geyerAnisotropicExchangeInteraction2024}. We anticipate that the gate fidelity can be further enhanced by optimal control techniques~\cite{rimbach-russSimpleFrameworkSystematic2023}.

\paragraph{Conclusion}
We have introduced the S2 qubit, a qubit encoding without requirements for complex control electronics, that can be controlled fully using only baseband pulses, possess no noticeably leakage states, and allows fast and high-fidelity two-qubit gates. Furthermore, the degenerate qubit states may allow for the implementation of holonomic quantum operations with even higher fidelity~\cite{zhangGeometricHolonomicQuantum2023,kolokProtocolsMeasureNonAbelian2024}. Our encoding fully leverages the potential of hole qubits and spin-orbit interaction through the g-tensor. Remarkably, our qubit can be implemented in state-of-the-art materials. This makes the S2 qubit a truly promising encoding to realize a large-scale quantum processor.

\paragraph{Acknowledgement}
We thank David DiVincenzo, Viatcheslav Dobrovitski, Edmondo Valvo, Menno Veldhorst, and  (in no particular order) for fruitful discussions. M.R-R. acknowledges support from NWO under Veni Grant (VI.Veni.212.223).
This research was partly supported by the EU through the H2024 QLSI2 project and partly sponsored by the Army Research Office under Award Number: W911NF-23-1-0110. The views and conclusions contained in this document are those of the authors and should not be interpreted as representing the official policies, either expressed or implied, of the Army Research Office or the U.S. Government. The U.S. Government is authorized to reproduce and distribute reprints for Government purposes notwithstanding any copyright notation herein.

\bibliography{lib}

\begin{thebibliography}{107}
\bibitem [{fid()}]{fidelity}%
  \BibitemOpen
  \href@noop {} {}\bibinfo {howpublished} {The average gate fidelity is defined as $F=\frac{1}{d(d+1)}(\text{Tr}[(U_\text{sim}U_\text{ideal}^\dagger)(U_\text{sim}U_\text{ideal}^\dagger)^\dagger] + |\text{Tr}[(U_\text{sim}U_\text{ideal}^\dagger)]|^2)$ , where $U_\text{sim}$ and $U_\text{ideal}$ are the simulated and ideal unitary matrix respectively and $d=2$ for single-qubit and $d=4$ for two-qubit operations}\BibitemShut {NoStop}%
\end{thebibliography}

\clearpage
\newpage
\mbox{~}

\onecolumngrid

\begin{center}
  \textbf{\large {Supplemental Material: \\ Exchange-Only Spin-Orbit Qubits in Silicon and Germanium}\\[.2cm]}
  Maximilian Rimbach-Russ, Valentin John, Barnaby van Straaten, and Stefano Bosco\\[.1cm]
  {\itshape QuTech, Delft University of Technology, Delft, The Netherlands}\\
\end{center}

\setcounter{equation}{0}
\setcounter{figure}{0}
\setcounter{table}{0}
\setcounter{section}{0}

\renewcommand{\theequation}{S\arabic{equation}}
\renewcommand{\thefigure}{S\arabic{figure}}
\renewcommand{\thesection}{S\arabic{section}}
\renewcommand{\bibnumfmt}[1]{[S#1]}
\renewcommand{\citenumfont}[1]{S#1}

\section*{abstract}
In this Supplemental Material, we derive the effective Hamiltonian of the spinless spin (S2) qubit and provide explicit expressions for all g-tensor components. Lastly, we report the complete numerical details and pulse shapes used for the numerical simulations of single qubit, readout, and two qubit gates.

\section{G-tensor for hole qubits in heterostructures}
Hole qubits in semiconductors and gate-defined quantum dots are well described by the Luttinger-Kohn-Bir-Pikus Hamiltonian. In the basis of total angular momentum eigenstates $\ket{j,m_j}=\lbrace\ket{\frac{3}{2},\frac{3}{2}},\ket{\frac{3}{2},-\frac{3}{2}},\ket{\frac{3}{2},\frac{1}{2}},\ket{\frac{3}{2},-\frac{1}{2}}\rbrace$ the Luttinger-Kohn-Bir-Pikus Hamiltonian reads~\cite{terrazosTheoryHolespinQubits2021}
\begin{align}
    H_\text{LKPB} = \left(\begin{array}{cccc}
        P+Q & 0 & S & R  \\ 
        0 & P+Q & R^\dagger & -S^\dagger  \\
        S^\dagger & R & P-Q & 0  \\
        R^\dagger & -S & 0 & P-Q  
    \end{array}\right).
    \label{eq:HLK}
\end{align}
The upper-left block $P+Q$ describe the energy of the $\pm\frac{3}{2}$ heavy-hole state, the lower-right block $P-Q$ describes the energy of the $\pm\frac{1}{2}$ light-hole state, $S$ describes the heavy-light-hole coupling with same spin, and $R$ describes the heavy-light-hole coupling with opposite spin direction. The operators are described by
\begin{align}
    P&= \frac{\hbar^2}{2m_0}\gamma_1 (k_x^2+k_y^2+k_z^2) -a_v (\varepsilon_{xx}+\varepsilon_{yy}+\varepsilon_{zz}),\\
    Q&= \frac{\hbar^2}{2m_0}\gamma_2 (k_x^2+k_y^2-2k_z^2)-\frac{b_v}{2} (\varepsilon_{xx}+\varepsilon_{yy}-2\varepsilon_{zz}),\\
    R&= \sqrt{3}\frac{\hbar^2}{2m_0}\left[-\gamma_2 (k_x^2-k_y^2)+i\gamma_3 k_x k_y+i\gamma_3 k_y k_x\right]+\frac{\sqrt{3}}{2}(\varepsilon_{xx}-\varepsilon_{yy})-id_v \varepsilon_{xy},\\
    S&= -\sqrt{3}\frac{\hbar^2}{2m_0}\gamma_3 \left[(k_x-i k_y)k_z+k_z(k_x-i k_y)\right]-d_v(\varepsilon_{xz}-\varepsilon_{yz}).
    \label{eq:LK}
\end{align}
Here, $p_\xi=\hbar k_\xi=-i \hbar \partial_\xi$ is the momentum operator in $\xi=x,y,z$ direction, $\hbar$ the reduced Planck constant, $m_0$ the bare electron mass, $\gamma_1$, $\gamma_2$, and $\gamma_3$ the material-dependent Luttinger parameters, and $a_v$, $b_v$, and $d_v$ the material dependent deformation parameters. The full system without magnetic field is consequently given by $H_\text{LKPB}+V$, where $V$ is the confinement potential. The impact of magnetic fields is through the Zeeman interaction terms and by substituting the momentum with the generalized momentum $\boldsymbol{p}\xrightarrow{} \boldsymbol{p}  + e\boldsymbol{A}$, where $\boldsymbol{A}$ is the electromagnetic vector potential. The Zeeman interaction term is $H_\text{Zeeman}=2\mu_{\rm B} \kappa\, \boldsymbol{J}\cdot\boldsymbol{B} + 2\mu_{\rm B} q (J_x^3 B_x + J_y^3 B_y + J_z^3 B_z)$, where $\mu_{\rm B}=e\hbar/(2m_0)$ is Bohr's magneton, $\hbar$ is the reduced Planck constant, $m_0$ is the electron mass, $e$ is the elementary charge, $\boldsymbol{B}=(B_x,B_y,B_z)^{\rm T}$ is the magnetic field, $\boldsymbol{J}=(J_x,J_y,J_z)^{\rm T}$ is the vector consisting of the spin-$\frac{3}{2}$ matrices, and $\kappa$ and $q$ are material dependent parameters. 

The g-tensor $\mathcal{G}$ of the hole qubit is then given by the linear response of the groundstate doublet of the full Hamiltonian $H_\text{LKPB}+V$ with respect to a magnetic field. We can then find $\mathcal{G}$ via perturbation theory following Ref.~\cite{venitucciSimpleModelElectrical2019}. For gate-defined quantum dots in silicon-germanium heterostructures grown in [001]-direction, the groundstate doublet is approximately a heavy-hole state and the light-hole sector is split by a gap $\Delta_\text{HL}$ that depends on the material and confinement~\cite{scappucciGermaniumQuantumInformation2020}. The low-energy Hamiltonian can then be approximated via block-diagonalization or Schrieffer-Wolff transformation~\cite{venitucciSimpleModelElectrical2019}
\begin{align}
    H_{\text{eff},ij} \approx H_{ij} -\sum_{k} \frac{\bra{\text{HH,0}}H_{ik}\ket{\text{LH,k}}\bra{\text{LH,k}}H_{kj}\ket{\text{HH,0}}}{E_{\text{LH},k}-E_\text{HH,0}}.
\end{align}
Here $E_{\text{LH},k}$ are the orbital energies, $E_{\text{HH},0}$ the ground state energy, and $k=(l,n)$ a multi-index that labels out-of-plane and in-plane wavefunctions. 
Considering strong confinement in z-direction $\Delta_\text{HL}\gg |E_{\text{LH},n}-E_\text{LH,0}|$, where $E_{\text{LH},n}$ are the energies from the in-plane orbital states, the sum over in-plane wavefunctions becomes trivial for most components due to $\sum_{n}\ket{\text{LH,n}}\bra{\text{LH,n}}\approx 1$. Collecting terms linear in $\boldsymbol{B}$, the g-tensor is then given by~\cite{venitucciSimpleModelElectrical2019,martinezHoleSpinManipulation2022a,abadillo-urielHoleSpinDrivingStrainInduced2023}
\begin{align}
    \mathcal{G}_{xx} &= 3q - \frac{6 \widetilde{\kappa} b_v (\braket{\epsilon_{xx}}-\braket{\epsilon_{yy}})}{\Delta_\text{HL}} - \frac{6 (\lambda \braket{p_x^2}-\lambda^\prime\braket{p_y^2})}{m_0 \Delta_\text{HL}}\\
    \mathcal{G}_{yy} &= -3q - \frac{6 \widetilde{\kappa} b_v (\braket{\epsilon_{xx}}-\braket{\epsilon_{yy}})}{\Delta_\text{HL}} + \frac{6 (\lambda \braket{p_y^2}-\lambda^\prime\braket{p_x^2})}{m_0 \Delta_\text{HL}},\\
    \mathcal{G}_{zz} &= 6\kappa -\frac{27}{2} q -2\gamma_h\\
    \mathcal{G}_{xy,yx} &= \pm  \frac{4\sqrt{3} \kappa d_v \epsilon_{xy}}{\Delta_\text{HL}} \mp \frac{12 \widetilde{\lambda} \braket{p_x p_y}}{m_0 \Delta_\text{HL}},
    \\ \mathcal{G}_{xz,yz} &=  \frac{4\sqrt{3} \kappa d_v \epsilon_{xz,yz}}{\Delta_\text{HL}} - \frac{(\mathcal{A}_{x,y}+\mathcal{B}_{x,y})\braket{p_{x,y}}}{m_0 \Delta_\text{HL}}.
\end{align}
Assuming that the groundstate wavefunction is separable in $z$ and $x,y$ direction, explicit expressions of the constants above are~\cite{michalLongitudinalTransverseElectric2021,abadillo-urielHoleSpinDrivingStrainInduced2023} 
\begin{align}
    \gamma_h &= \frac{6\gamma_3^2\hbar}{m_0}\sum_{l} \frac{\left|\bra{\text{HH}_0}k_z\ket{\text{LH}_l}\right|^2}{E_{\text{LH},l}-E_{\text{HH},0}}\\
    \eta_h &= -\Delta_\text{LH}\sum_{l} \frac{2\text{Im}\left(\bra{\text{HH}_0}z\ket{\text{LH}_l}\bra{\text{LH}_l}k_z\ket{\text{HH}_0}\right)}{E_{\text{LH},l}-E_{\text{HH},0}}\\
    \widetilde{\eta}_h &= \Delta_\text{LH}\sum_{l} \frac{\text{Im}\left(\bra{\text{HH}_0}k_zz+zk_z\ket{\text{LH}_l}\braket{\text{LH}_l|\text{HH}_0}\right)}{E_{\text{LH},l}-E_{\text{HH},0}}\\
    \widetilde{\kappa} &=\kappa - \widetilde{\eta}_h.
\end{align}
Terms that require a non-separability of the groundstate wavefunction between $z$ and $x,y$ direction are~\cite{martinezHoleSpinManipulation2022a}
\begin{align}
    \mathcal{A}_{x,y} &= 6\kappa\Delta_\text{LH}\sum_{l} \frac{\text{Re}\left(\bra{\text{HH}_0}p_{x,y}p_z\ket{\text{LH}_l}\braket{\text{LH}_l|\text{HH}_0}\right)}{E_{\text{LH},l}-E_{\text{HH},0}}\label{eq:nonsepA}\\
    \mathcal{B}_{x,y} &= 3\gamma_3^2\Delta_\text{LH}\sum_{l} \frac{\text{Im}\left(\bra{\text{HH}_0}k_z z+k_z z\ket{\text{LH}_l}\bra{\text{LH}_l}p_{x,y}p_z\ket{\text{HH}_0}\right)}{E_{\text{LH},l}-E_{\text{HH},0}}.\label{eq:nonsepB}
\end{align}
\clearpage

\section{Voltage susceptibility}
We have performed Schrödinger-Poisson simulations of a gate layout similar to Fig.~1a) and a heterostructure as discussed in Ref.~\cite{wangOperatingSemiconductorQuantum2024a} to estimate the tunability of our S2 qubit.
\begin{figure}[h!]
    \centering
    \includegraphics[width=0.8\columnwidth]{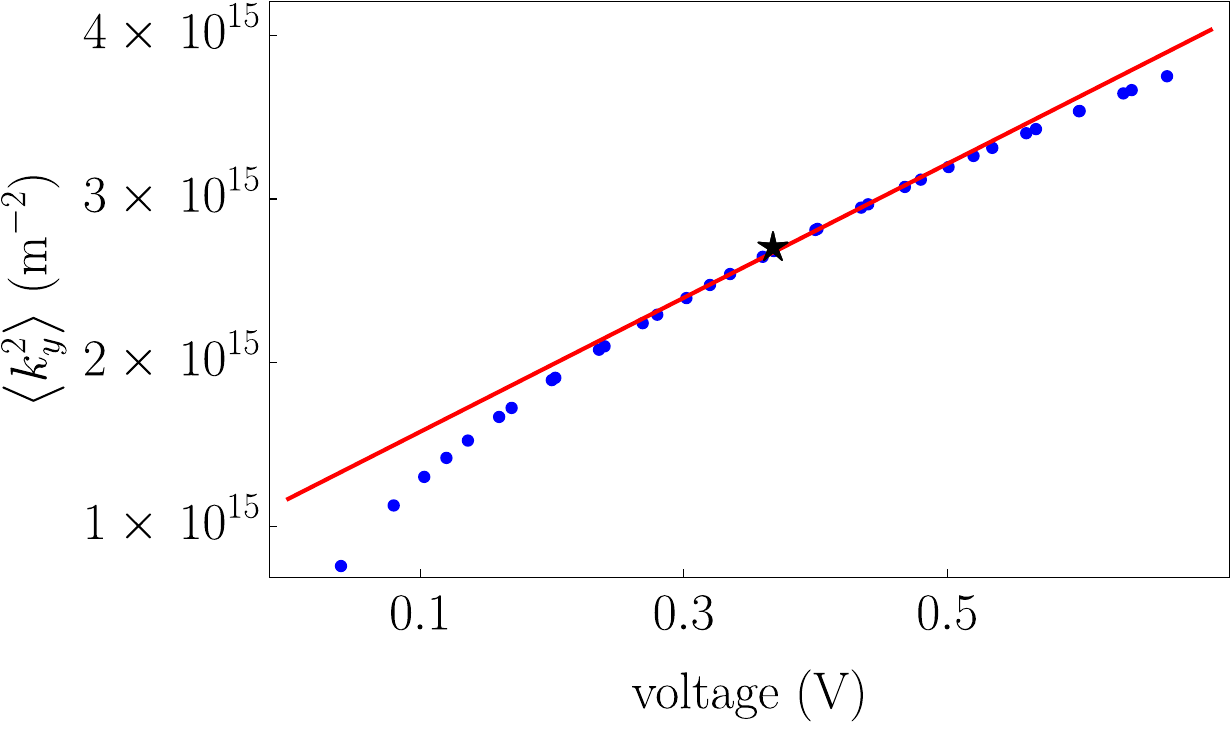}
    \caption{Expectation value $\braket{k_y^2}$ using a gate layout similar to Fig.~1a) in the main text as a function of voltage applied to the center gate of radius $\unit[80]{nm}$. Blue dots are extracted simulation, the star ($\star$) highlights the S2 operation regime, and the solid line is linear regression with $\braket{k_y^2}=\unit[1.2\times 10^{15}]{m^{-2}}+\unit[4.1\times 10^{15}]{m^{-2}V^{-1}}\,V_g$ around the operation point. Heterostructure and device layout are taken from Ref.~\cite{wangOperatingSemiconductorQuantum2024a} and material parameters Ref.~\cite{terrazosTheoryHolespinQubits2021}. Schrödinger-Poisson simulations are performed using QTCAD~\cite{philippopoulosAnalysis3DTCAD2024}.}
    \label{fig:supp_1}
\end{figure}

\section{Simulation of qubit dynamics}
For all simulations of the qubit dynamic, we solve the time-dependent Schr{\"o}dinger equation 
\begin{align}
    i\hbar\frac{d}{dt} \ket{\psi(t)} = H \ket{\psi(t)}
\end{align}
for a full set of basis states and using the \texttt{NDSOLVE} command of Mathematica with sufficiently high precision. The time evolution operator can then be constructed from the final state vectors.

\subsection{Single-qubit operations}
The single-qubit dynamics is modelled by the Hamiltonian
\begin{align}
    H_{1q} &= b_x(t)  \sigma_x + b_y(t) \sigma_y + b_z \sigma_z.
    \label{eq:QubitHam_supp}
\end{align}
Here, $\sigma_x$, $\sigma_y$, and $\sigma_z$ are the Pauli matrices and $b_{x,y}(t)$ are the (time-dependent) control parameters. 
We consider for simplicity a constant pulse amplitude $b_x(t)=b_x$ and constant calibration errors $b_y(t)=b_y$ and $b_z(t)=b_z$. We further assume that the g-tensor is a linear function of voltage, $b_x\propto V_\text{g}$. However, we note that this is not an requirement. 
The the three-pulse CORPSE sequence is then given by
\begin{align}
b_x(t)=
\begin{cases} 
      b_x &  t\in[ t_0,t_1) \\
      -b_x &  t\in[ t_1,t_2) \\
      b_x &  t\in[ t_2,t_3) \\
      0 &  \text{otherwise} \\
   \end{cases}
\end{align}
with the pulse times for the short-CORPSE  
\begin{align}
	b_x (t_0-t_1) &=(\theta/2 - \arcsin(\theta/2))/2\\
	b_x (t_1-t_2) &=(2\pi - 2\arcsin(\theta/2))/2\\
	b_x (t_2-t_3) &=(\theta/2 - \arcsin(\theta/2))/2
\end{align}
and for the conventional CORPSE composite pulse sequence
\begin{align}
	b_x (t_0-t_1) &=(2\pi +\theta/2 - \arcsin(\theta/2)/2)\\
	b_x (t_1-t_2) &=(2\pi - \arcsin(\theta/2)/2)\\
	b_x (t_2-t_3) &=(\theta/2 - \arcsin(\theta/2))/2.
\end{align}
\subsection{Readout}
For readout, we use the Hamiltonian
\begin{align}
    H=& H_{1q}^{(A)}+H_{1q}^{(B)} + \varepsilon(t) \ket{S_{02}}\bra{S_{02}}+\sqrt{2}t_c\cos(\theta_\text{SOI})\ket{S}\bra{S_{02}} \nonumber\\
    &+i\sqrt{2}t_c(t)\sin(\theta_\text{SOI})\,\boldsymbol{n}\cdot\ket{\boldsymbol{T}}\bra{S_{02}} +\text{h.c.},
\end{align}
where $H_{1q}^{(A,B)}$ is the single-qubit Hamiltonian~\eqref{eq:QubitHam_supp} of qubit $A,B$, $\boldsymbol{n}$ is the normalized spin-orbit vector, $\theta_\text{SOI}$ the spin-flip angle, and $\boldsymbol{T}_{AB}=(T_x,T_y,T_z)^T$ is a vector consisting of the spin-1 triplet states with $\ket{T_{x,y}}=i^{1/2\mp 1/2}(\ket{00}\mp\ket{11})/\sqrt{2}$ and $\ket{T_z}=(\ket{01}+\ket{10})/\sqrt{2}$. Since the tunnel coupling is approximately given by the overlaps of the wavefunctions between qubit A and B, we model the tunnel coupling as a Gaussian function of voltage~\cite{burkardCoupledQuantumDots1999}. The explicit simulation parameters for Fig.~3b) in the main text are
\begin{align}
	b_x^{(A)}(t) &= \unit[0.3]{GHz}\begin{cases} 
      0 &  t<0 \\
      \frac{t}{t_r} &  t\in[ 0,t_r] \\
      1 &  \text{otherwise} \\
   \end{cases}\\
   b_x^{(B)}(t) &= \unit[0.4]{GHz}\begin{cases} 
      0 &  t<0 \\
      \frac{t}{t_r} &  t\in[ 0,t_r] \\
      1 &  \text{otherwise} \\
   \end{cases}\\
   b_y^{(A)}(t) &= \unit[0.1]{GHz}\begin{cases} 
      0 &  t<t_r/2 \\
      \frac{2t-t_r}{t_r} &  t\in[ t_r/2,t_r] \\
      1 &  \text{otherwise} \\
   \end{cases}\\
   b_y^{(B)}(t) &= -\unit[0.1]{GHz}\begin{cases} 
      0 &  t<t_r/2 \\
      \frac{2t-t_r}{t_r} &  t\in[ t_r/2,t_r] \\
      1 &  \text{otherwise} \\
   \end{cases}\\
   t_c(t) &= \unit[3]{GHz}\begin{cases} 
      0 &  t<0 \\
      \frac{\exp\left[\log(10)\left(\frac{t}{t_r}\right)^2\right] - 1}{t_r-1} &  t\in[0,t_r] \\
      1 &  \text{otherwise} \\
   \end{cases}\\
   \varepsilon(t) &= \unit[0.1]{meV}\begin{cases} 
      1 &  t<0 \\
      1-\frac{2t}{t_\text{PSB}} & \text{otherwise}
   \end{cases}.
\end{align}
Here, $t_r=\unit[10]{ns}$ is the ramp time to elongate the wavefunction and $t_\text{PSB}=\unit[86]{ns}$ the total time for the readout. 
\subsubsection{Two-qubit gates}
For two-qubit gate simulations, we use the Hamiltonian
\begin{align}
    H_{2q}= H_{1q}^{(A)}+H_{1q}^{(B)} + \frac{J(t)}{4} \boldsymbol{\sigma}^{(A)}\cdot R_{\boldsymbol{n}} (2\theta_\text{SOI})\boldsymbol{\sigma}^{(B)}
\end{align}
where $J$ is the electrically tunable exchange coupling and $R_{\boldsymbol{n}} (2\theta_\text{SOI})$ is a 3D rotation matrix originating from the influence of the spin-orbit interaction~\cite{geyerAnisotropicExchangeInteraction2024}. Since $J\propto t_c^2$, we also model $J(t)$ assuming a Gaussian profile. The explicit simulation parameters for Fig.~3c) in the main text are
\begin{align}
	b_x^{(A)}(t) &= \unit[0.3]{GHz}\begin{cases} 
      \frac{t}{t_r} &  t\in[ 0,t_r) \\
      1 &  t\in[ t_r,2t_r) \\
      \frac{3t_r-t}{t_r} &  t\in[ 2t_r,3t_r) \\
      0 &  \text{otherwise} \\
   \end{cases}\\
   b_x^{(B)}(t) &= \unit[0.4]{GHz}\begin{cases} 
      \frac{t}{t_r} &  t\in[ 0,t_r) \\
      1 &  t\in[t_r,2t_r) \\
      \frac{3t_r-t}{t_r} &  t\in[ 2t_r,3t_r) \\
      0 &  \text{otherwise} \\
   \end{cases}\\
   b_y^{(A)}(t) &= 0\\
   b_y^{(B)}(t) &= 0\\
   J(t) &= J_0\begin{cases} 
      \frac{\exp\left[\log(10)\left(\frac{t}{t_r}\right)^2\right] - 1}{t_r-1} &  t\in[0,t_r) \\
      \frac{9}{t_r-1} &  t\in[ t_r,2t_r) \\
      \frac{\exp\left[\log(10)\left(\frac{3t_r-t}{t_r}\right)^2\right] - 1}{t_r-1} &  t\in[ 2t_r,3t_r] \\
      0 &  \text{otherwise}, \\
   \end{cases}
\end{align} 
where $t_3=\unit[18]{ns}$ and $J_0$ is chosen such that $\int_{0}^{3t_r}J(t)(\cos(2 \theta) (\boldsymbol{n}_2^2 + \boldsymbol{n}_3^2) + \boldsymbol{n}_1^2)=\pi$ with $\boldsymbol{n}=(\cos(\phi),\sin(\phi),0)^T$.

\end{document}